\documentclass[11pt]{article}
\pdfoutput=1
\usepackage{jheppub}
\usepackage{tabu}
\usepackage[vcentermath]{youngtab}
\usepackage[usenames,dvipsnames,table]{xcolor}
\usepackage{graphicx,amsmath,amssymb,amsthm,multirow,array,bm,bbm,esint}
\usepackage[mathscr]{eucal}
\usepackage[bbgreekl]{mathbbol}
\usepackage{epsf,amsfonts}
\usepackage{slashed}
\usepackage[numbers,sort&compress]{natbib}
\usepackage{minibox}
\usepackage{rotating}
\usepackage{pdflscape}
\usepackage{array,tikz-cd}
\usepackage{subfigure}

\usepackage{slashed}

\definecolor{rust}{rgb}{0.8,0.2,0.2}

\newcommand{\brk}[1]{\left [ #1 \right ]}

\newcommand{\half}{\frac{1}{2}}

\newcommand{\QSK}{\mathcal{Q}_{_{SK}}}
\newcommand{\QSKb}{\overline{\mathcal{Q}}_{_{SK}}}
\newcommand{\QKMS}{\mathcal{Q}_{_{KMS}}}
\newcommand{\QKMSb}{\overline{\mathcal{Q}}_{_{KMS}}}

\newcommand{\Qzero}{\mathcal{Q}^0_{_{KMS}}}
\newcommand{\Qbeta}{\mathscr{L}_{_{KMS}}}

\newcommand{\Q}{\mathcal{Q}}
\newcommand{\Qb}{\overline{\mathcal{Q}}}

\newcommand{\Op}[1]{\mathbb{#1}}

\newcommand{\SKAv}[1]{\mathbb{#1}_{{av}}}

\newcommand{\SKDif}[1]{\mathbb{#1}_{{dif}}}

\newcommand{\SKAdv}[1]{\mathbb{#1}_{adv}}
\newcommand{\SKRet}[1]{\mathbb{#1}_{ret}}

\newcommand{\SKG}[1]{\mathbb{#1}_{_G}}
\newcommand{\SKGb}[1]{\mathbb{#1}_{_{\overline{G}}}}

\newcommand{\gradcomm}[2]{ \brk{ #1, #2 }_{\scriptscriptstyle \pm} }

\newcommand{\thb}{{\bar{\theta}} }

\newcommand{\SF}[1]{\mathring{#1}}
\newcommand{\Dsf}{\SF{\mathcal D}}

\newcommand{\Ascr}{\mathscr{A}}

\newcommand{\Fs}{\SF{\mathscr{F}}}

\newcommand{\deltaB}{\delta_{_ {\bm\beta}}}

\newcommand{\delKMS}{{\Delta}_{_ {\bm\beta}} }

\newcommand{\Kref}{{\bm \beta}}
\newcommand{\Lref}{\Lambda_\Kref}

\newcommand{\gref}{{\sf g}}

\newcommand{\etaref}{\bm \eta}

\makeatletter
  \newcommand\Ttiny{\@setfontsize\Ttiny{1pt}{2}}
\makeatother

\newcommand{\Cref}{{\sf C}}

\newcommand{\lieD}{\pounds}

\newcommand{\Kbeta}{{\bm{\beta}}}
\newcommand{\LambdaB}{\Lambda_{\bm{\beta}}}

\newcommand{\hfields}{{\bm \Psi}}

\newcommand{\JH}{\mathrm{J}_{H}}

\newcommand{\THall}{\mathrm{T}_{H}}

\newcommand{\smallT}{{\sf \!{\scriptscriptstyle{T}}}}

\newcommand{\UT}{U(1)_{\scriptstyle{\sf T}}}
\newcommand{\AT}{{\sf A^{ \!{\scriptscriptstyle{(T)}}}}\!}

\newcommand{\PS}{{\rm H}_S}
\newcommand{\PV}{{\rm H}_V}

\newcommand{\LS}{{\overline{\rm H}}_S}
\newcommand{\GV}{{\overline{\rm H}}_V}

\newcommand{\LT}{{\rm L}_{\,\smallT}}

\def\ecite#1{$_\text{\scriptsize{\cite{#1}}}$}

\title{Two roads to hydrodynamic effective actions: a comparison}

\author[a]{Felix M. Haehl}
\author[b]{\!, R.\ Loganayagam}
\author[c]{\!, Mukund Rangamani}
\affiliation[\,a]{Department of Physics and Astronomy, University of British Columbia,\\
6224 Agricultural Road, Vancouver, B.C.\ V6T 1Z1, Canada.}
\affiliation[\,b]{International Centre for Theoretical Sciences (ICTS-TIFR), \\
Shivakote, Hesaraghatta Hobli, Bengaluru 560089, India.}
\affiliation[\,c]{
Center for Quantum Mathematics and Physics (QMAP)  \\
Department of Physics, University of California, Davis, CA 95616 USA.}
\emailAdd{f.m.haehl@gmail.com}
\emailAdd{nayagam@gmail.com}
\emailAdd{mukund@physics.ucdavis.edu}

\abstract{
We make a detailed comparison between two attempts in recent years to construct hydrodynamic effective actions: we compare our work \cite{Haehl:2013hoa,Haehl:2014zda,Haehl:2015pja,Haehl:2015foa,Haehl:2015uoc,Haehl:2016pec,Haehl:2016uah} with that of Crossley-Glorioso-Liu \cite{Crossley:2015evo} and Glorioso-Liu \cite{Glorioso:2016gsa}. The general philosophy espoused by the two approaches has a degree of overlap, despite various differences. We will try to outline the similarities to eke out the general lessons that have been uncovered, hoping that it will ease the access to the subject for interested readers.
}

\begin{document}
\maketitle


\section{Hydrodynamic actions in the current decade}
\label{sec:intro}

In the past few years we have written several papers outlining a general framework for the effective low energy field theory in local equilibrium \cite{Haehl:2013hoa,Haehl:2014zda,Haehl:2015pja,Haehl:2015foa,Haehl:2015uoc,Haehl:2016pec,Haehl:2016uah}. In the same period, Crossley-Glorioso-Liu \cite{Crossley:2015evo} and Glorioso-Liu \cite{Glorioso:2016gsa} have written two papers on the subject which have a reasonable amount of overlap with our considerations. They also wrote an earlier paper \cite{Crossley:2015tka} which attempts to make contact with the gravitational description of fluids a la fluid/gravity \cite{Bhattacharyya:2008jc,Hubeny:2011hd}, with partial success. A similar analysis from the gravity side was also carried out concurrently in \cite{deBoer:2015ija}. The following is an attempt to  compare the two approaches by us (HLR) and that by the MIT groups (CGL and GL, respectively).

 Much of our analysis builds on our first work together on anomalous effective actions \cite{Haehl:2013hoa}. This analysis which focused on flavour anomaly contributions, was the first to reveal interesting surprises about what we should expect for hydrodynamic effective field theories. Two main physical issues were made apparent:
an intrinsic doubling of degrees of freedom (even in the absence of dissipation)\footnote{ It would be fair to say that the standard lore till this publication would have suggested that non-dissipative transport would be obtained from a conventional action principle without any doubling. This intuition is spectacularly untrue, as we now know based on the analysis of  \cite{Haehl:2013hoa,Haehl:2014zda,Haehl:2015pja}.} which was reminiscent of the Schwinger-Keldysh construction, and a rigid constraint on couplings between these copies. The field theory analysis also gave a nice heuristic picture in the gravitational context, as explained in Fig.~1 of \cite{Haehl:2013hoa}.\footnote{ A similar picture was subsequently advocated in \cite{Crossley:2015tka}.} 

It is worth emphasizing the earlier work of Nickel and Son \cite{Nickel:2010pr} which suggested an interesting way to write down hydrodynamic variables inspired by holographic considerations. Likewise, the work of Dubovsky et.al.\ \cite{Dubovsky:2011sk}, was the first to construct anomalous effective actions, for abelian flavour anomalies in two dimensions. This turns out to be a special case, where one can eschew the doubling. The analysis carried out in \cite{Haehl:2013hoa} uses the hydrodynamic variables described in \cite{Dubovsky:2011sk,Dubovsky:2011sj}, which have also been examined systematically for action principles in the non-dissipative sector in \cite{Nicolis:2011ey,Bhattacharya:2012zx,Haehl:2013kra,Geracie:2014iva}. The way of thinking about the hydrodynamic fields as living on the black hole horizon and viewing the maps therefrom to the boundary is inspired in large part by \cite{Nickel:2010pr}. 

Whilst the anomaly story was interesting, it was clear from the rich phenomenology of hydrodynamics that there was more structure to be unearthed. Naive implementations of the formalism failed to capture Lorentz anomalies (mixed flavour/gravitational anomalies), and there were puzzles about the inclusion of some known forms of non-dissipative transport such as Hall viscosity \cite{Nicolis:2011ey,Bhattacharya:2012zx,Haehl:2013kra,Geracie:2014iva}. These issues therefore spurred us to consider the following question:
\emph{Can we, given the basic axioms of hydrodynamics as the theory of conserved currents, subject to the local form of second law of thermodynamics, give a complete classification of the structural form of hydrodynamic currents?}

This question was answered by us in the affirmative in \cite{Haehl:2015pja} (a brief summary of results appears in \cite{Haehl:2014zda}). The basic theorem proved here states that hydrodynamic transport has an eightfold classification, with seven distinct adiabatic classes and a single class of dissipative transport.\footnote{ The classes were named A (anomalies), B (Berry-like), C (conserved entropy), D(dissipative), $\PS$ (hydrostatic scalars), $\LS$ (hydrodynamic scalars), $\PV$ (hydrostatic vectors), and $\GV$ (hydrodynamic vectors).} The proof was essentially constructive in the sense that it gave explicitly the general form of all possible constitutive relations $\{T^{\mu\nu},J^\mu,J^\mu_S\}$ consistent with the phenomenological axioms of hydrodynamics. The latter state that the dynamics of the fluid is encoded in conservation laws for the stress tensor $T^{\mu\nu}$ and flavour current $J^\mu$, subject to the existence of an entropy current $J^\mu_S$, which satisfies a local version of the second law:
\begin{equation}\label{eq:hydroBasics}
  \nabla_\nu T^{\mu\nu} = J_\nu \cdot F^{\mu\nu} + \THall^{\mu\perp} \,, \qquad
  D_\nu J^\nu = \JH^\perp \,,\qquad
  \nabla_\mu J_S^\mu \geq 0 \,,
\end{equation}
where Hall currents (with subscript $H$) denote the non-conservation due to anomalies. 
Our classification of solutions to this set of equations relied heavily on the intuition derived from the hydrostatic partition function analysis of \cite{Banerjee:2012iz,Jensen:2012jh}, and more crucially on  some the results derived by S. Bhattacharyya in a series of papers \cite{Bhattacharyya:2013lha,Bhattacharyya:2014bha}.\footnote{ An important precursor work which proved influential in these developments was the first comprehensive analysis of second law constraints on hydrodynamic transport \cite{Bhattacharyya:2012nq} (see \cite{Romatschke:2009kr} for earlier results). } 

A few  things became apparent in the discussion of the eightfold classification which are worth highlighting:
\begin{itemize}
\item The space of adiabatic transport is far richer than hitherto appreciated (it makes up 7 of the 8 classes). The word adiabatic connotes in a technical sense the off-shell conservation of the entropy current $J^\mu_S$ in \eqref{eq:hydroBasics}. Thus adiabatic transport here refers to transport with vanishing entropy production.
\item The choice of hydrodynamic variables is best encoded in a thermal vector $\Kbeta^\mu = u^\mu/T$ (which encodes fluid velocity and temperature) and a thermal twist $\LambdaB = \frac{\mu}{T} - \Kref^\mu A_\mu$ (which encodes the flavour chemical potential).\footnote{ This was very much inspired by the structure of hydrostatic equilibrium as described initially in \cite{Banerjee:2012iz,Jensen:2012jh} and analysis of anomalous transport in that formalism \cite{Jensen:2012kj,Jensen:2013kka,Jensen:2013rga}.}  
\item A naive single copy effective action (of a Landau-Ginzburg form) only captures 2 of the 7 adiabatic classes. In this sector entropy current is a Noether current associated with thermal translations/gauge transformations along $\{\Kbeta^\mu,\LambdaB\}$.
\item The Landau-Ginzburg action for two of the classes (called $\PS$ and $\LS$) can be viewed as the worldvolume action of a space filling fluid/Brownian brane. The basic degrees of freedom are the maps from a reference worldvolume equipped with a rigid thermal vector ${\mathbb \bbbeta}^a$ and twist $\Lambda_{{\mathbb \bbbeta}}$, which get pushed-forward to the physical fluid spacetime under the embedding map. 
\end{itemize} 

Given our experience with anomalies, it was apparent that one should attempt to write down an effective action which involves two sets of degrees of freedom to obtain the remaining 5 adiabatic and 1 dissipative sector. Focusing on the adiabatic sector, we argued in \cite{Haehl:2014zda,Haehl:2015pja} that:
\begin{itemize}
\item We should look for a formulation that takes the Schwinger-Keldysh construction seriously, as presaged in \cite{Haehl:2013hoa}. As a first step in the effective theory of hydrodynamics, this means a doubling of degrees of freedom.  
\item A naive implementation of Schwinger-Keldysh however would be in tension with microscopic unitarity, for it would involve couplings between the two sets of degrees of freedom. These \emph{influence functionals} in the language of \cite{Feynman:1963fq} ought to be constrained by demanding that they respect the unitarity constraints, and also respect the thermal KMS constraints (suitably applied to near-equilibrium for almost Gibbsian density matrices).\footnote{ The reader is invited to consult Sections 13  and 18 of \cite{Haehl:2015pja} for our detailed reasoning.} 
\item We found empirically that these constraints would be upheld in the hydrodynamic effective action, should we postulate an  emergent low energy symmetry, called the thermal KMS $\UT$ symmetry, to constrain the set of low energy couplings \cite{Haehl:2014zda,Haehl:2015pja}. The transformation of fields under this symmetry can roughly be described as follows:\footnote{ For details, we refer to \cite{Haehl:2015pja,Haehl:2015uoc}.}
\begin{enumerate}
 \item On `retarded' fields: a thermal diffeomorphism/gauge transformation along $\{\Kbeta^\mu,\LambdaB\}$
 \item On `advanced' fields: a similar thermal shift by their associated `retarded' partner
\end{enumerate}
\item With this inspired guess we were able to construct an adiabatic action for the 7 classes of transport (called Class $\LT$ in \cite{Haehl:2015pja}). 
Whilst no derivation was provided for the $\UT$ symmetry, we showed that 
\begin{enumerate}
\item The symmetry closes on the set of low energy fields (satisfies desired Wess-Zumino conditions).
\item Entropy is a Noether current for this $\UT$ symmetry, and is conserved due to our restriction to the adiabatic sector.
\end{enumerate}
\end{itemize}

The hydrodynamic effective Lagrangian took the following simple form:
\begin{equation}
\mathcal{L}_\smallT = N^\mu[\hfields] \, \AT_\mu + \frac{1}{2} T^{\mu\nu}[\hfields] \,\tilde{g}_{\mu\nu} + J^\mu[\hfields] \cdot \tilde{A}_\mu
\label{eq:LTact}
\end{equation}	
with $\AT_\mu$ being the gauge potential for the $\UT$ symmetry, $\tilde{g}_{\mu\nu}$ and $\tilde{A}_\mu$ being the Schwinger-Keldysh inspired partners for the background metric $g_{\mu\nu}$ and gauge field $A_\mu$ sources. The collection of hydrodynamic fields and sources is $\hfields = \{\Kbeta^\mu,\LambdaB;g_{\mu\nu},A_\mu\}$. Further, 
\begin{equation}
N^\mu = J^\mu_S + \Kbeta_\nu T^{\mu\nu} + (\LambdaB+\Kbeta^\nu A_\nu) \cdot J^\mu \,, 
\label{eq:Ncurrent}
\end{equation}	
is the Noether current for thermal translations (also often referred to as non-canonical part of the entropy current, or as free energy current up to a rescaling by temperature). The Lagrangian \eqref{eq:LTact} summarizes the entire adiabatic sector of our classification of transport in a simple and efficient way: {\it a given set of currents $\{N^\mu,T^{\mu\nu},J^\mu\}$ describes adiabatic transport if and only if the associated action \eqref{eq:LTact} is invariant under $\UT$.} According to the standard rules of effective field theory, this statement proves that at least for adiabatic hydrodynamics $\UT$ exists as an emergent symmetry. 

Following the publication of the eightfold classification, the two papers \cite{Crossley:2015tka,deBoer:2015ija} appeared, which take further steps towards a holographic realization of related ideas. Both of these works attempt to generalize \cite{Nickel:2010pr} and get the ideal fluid part of the action (which is equivalent to the hydrostatic partition function to leading order in gradients) accurately. The group CGL \cite{Crossley:2015tka} also attempted to get the second order hydrostatic partition function, but for a reason unknown to us encountered obstructions.  To be specific, the hydrostatic partition function for a neutral fluid at second order in gradients comprises of three potential contributions. Only two were recovered from the holographic analysis --  we recall that the hydrostatic partition function  has three second order contributions, cf.,\footnote{ We will refer to equation numbers of papers mentioned in the text with suitable subscripts to indicate the reference in question.} Eq.~(1.6)\ecite{Banerjee:2012iz} -- the term proportional to fluid vorticity squared was not obtained.
Eq.~(14.35)\ecite{Haehl:2015pja} gives the correct expression for the hydrostatic partition function  that reproduces the transport coefficients derived from holography.

Given the success in explaining the phenomenological axioms of adiabatic hydrodynamics under the assertion of an emergent $\UT$ symmetry via \eqref{eq:LTact}, for us the key question was how to justify $\UT$ in a quantum field theoretic way and whether this would suffice to capture the dissipative sector. Earlier work of Kovtun et.al.\ \cite{Kovtun:2014hpa} whose dissipative action was broadly speaking of the form \eqref{eq:LTact} (without $\UT$ gauge field) lent strength to our conjecture. Their analysis was reminiscent of older work on dissipative systems in statistical mechanics \cite{Martin:1973zz} and stochastic dynamics \cite{Parisi:1982ud}. The effective actions derived by these papers incorporates a topological (BRST)  supersymmetry. 
This brought these set of questions close to the studies of topological field theory  defined over the moduli space of solutions of certain equations. The work by Vafa-Witten \cite{Vafa:1994tf} over the moduli space of instantons was paradigmatic in this regard and so was the subsequent explication by Dijkgraaf -Moore \cite{Dijkgraaf:1996tz} and others \cite{Blau:1996bx}  on the $N_\smallT=2$ structure that correctly counts the solutions without signs.
Understanding this proved key to our subsequent explorations. 

Inspired by these statements, we outlined a general philosophy, which takes microscopic unitarity and KMS condition as the central guiding principles for deriving the correct low energy symmetries governing hydrodynamic transport. This was explained in two talks at Strings 2015 \cite{Rangamani:2015str,Loganayagam:2015str}. The basic ideas were explained in broad terms, laying out a framework for the construction of hydrodynamic effective actions in the Fluid Manifesto  \cite{Haehl:2015foa}. The fundamental statement was that there should exist an $\mathcal{N}_\smallT = 2$ equivariant supersymmetry algebra of thermal translations, which completely determines consistent hydrodynamic transport. 

Shortly afterward we received \cite{Crossley:2015evo}, whose intent was to construct actions for dissipative hydrodynamics; we described our version of an action principle in  \cite{Haehl:2015uoc} closely afterward. The two papers identify a set of symmetries which should govern the low energy dynamics. The thrust of the two approaches appears to be rather different in their phrasing and notation, but there are more similarities than superficially visible. To be clear, our \cite{Haehl:2015uoc} follows strongly on the lessons learnt in the eightfold classification in \cite{Haehl:2015pja}. A first principles explanation of some of the statements  and underlying symmetries was provided more recently in \cite{Haehl:2016pec,Haehl:2016uah}. A short while ago we received a second paper from the GL,  linking entropy production with microscopic unitarity \cite{Glorioso:2016gsa}. The hydrodynamic action of this paper has been translated into a form that makes comparison more seamless; indeed we shall see that the current picture hews more closely to the philosophy of \cite{Haehl:2015pja,Haehl:2015uoc}. We hope that the current discussion will serve to bring out some of  these similarities.

\noindent 
\emph{Note added:} As this note was nearing completion, we became aware of further work on the subject: the ideas of CGL are further developed in \cite{Gao:2017bqf,Glorioso:2017fpd}, and \cite{Jensen:2017kzi} also deals with the same problem by taking some inspiration from both our formalism and that of CGL.

\section{Effective actions for hydrodynamics}
\label{sec:paper1}

Let us take stock of the constructions in the two approaches, comparing first the basic variables, the symmetries, and the form of the action. 

\subsection{Hydrodynamic variables}

 The standard current algebra formulation of hydrodynamics, by which we mean the process whereby we write down the constitutive relations $\{T^{\mu\nu},J^\mu,J^\mu_S\}$ in terms of the hydrodynamic variables and background sources, identifies as variables the fluid velocity and intensive parameters corresponding to local thermal equilibrium.  While early discussions on the subject \cite{Nickel:2010pr,Dubovsky:2011sk,Dubovsky:2011sj,Bhattacharya:2012zx,Haehl:2013kra,Geracie:2014iva,Haehl:2013hoa} attempted to use Lagrangian fluid variables, it was found in \cite{Haehl:2015foa} that these variables do not work well outside the entropy frame. In particular, the Lagrangian variables of \cite{Dubovsky:2011sj}, which are best viewed as spatial positions of fluid cells at an instant of time, involve an additional volume preserving diffeomorphism symmetry, which one views as responsible for entropy conservation. While this is true for part of the adiabatic sector, there exist adiabatic fluid constitutive relations wherein there is a non-trivial entropy current, which nevertheless leads to no dissipation.\footnote{ This issue was of import in addressing the contribution of mixed flavour/gravitational anomalies to transport.}

The more natural variables for the current algebraic formulation of hydrodynamics, are the thermal vector and twist $\Kbeta^\mu$ and $\LambdaB$ introduced in \cite{Haehl:2015pja}. These allow for a covariant presentations. One can show that the Lagrangian variables alluded to above, can be obtained by a suitable gauge fixing -- the explicit argument can be found in Appendix  B of \cite{Haehl:2015pja}.

The true dynamical degrees of freedom for an off-shell effective action are not the entirety of the thermal vector and twist, but the part of them which is orthogonal to a constraint surface. These physical modes turn out to be   Goldstone modes $X^\mu(\sigma)$ and $c(\sigma)$ which are maps from some reference worldvolume geometry onto the physical fluid. Put simply, the fluid dynamical effective field theory, is a sigma model of a space filling brane, with the physical fluid thermal vector and twist being push-forwards of a reference thermal vector $\Kref^a$ and twist $\Lref$ 
that live on a fiducial worldvolume (with intrinsic coordinates $\sigma^a$). Initially we viewed the maps as running from the physical spacetime to the reference configuration; cf., Eq.~(7.5)\ecite{Haehl:2015pja} and (7.6)\ecite{Haehl:2015pja} (see Fig.~3 of \cite{Haehl:2015pja} which also uses blackboard bold font to pick out reference fields, e.g., ${\mathbb \bbbeta}^a$ and twist $\Lambda_{{\mathbb \bbbeta}}$). In our subsequent works we realized that the sigma model interpretation is more natural, which prompts the introduction of $X^\mu, c$ variables introduced in Section 3 of \cite{Haehl:2015uoc}; this rationale is explained explicitly in Footnote 7 of \cite{Haehl:2015uoc}.

In contrast, the authors of \cite{Crossley:2015evo} continue to use a variant of the Lagrangian variables
in the following sense. They recognize that the Goldstone modes $X^\mu(\sigma), c(\sigma)$ are the physical degrees of freedom. However, rather than pick a reference thermal vector and twist on the worldvolume, they choose to work in a non-covariant spacetime where the background Lorentz structure has been deformed to  allow spatial diffeomorphisms and  space dependent time reparametrizations (cf., Eqs.~(1.27)\ecite{Crossley:2015evo} and (1.28)\ecite{Crossley:2015evo}). This is akin to the story of foliation preserving diffeomorphisms described in \cite{Horava:2009uw}. In addition they identify the fluid variables in terms of the Goldstone modes in Eq.~(5.1)\ecite{Crossley:2015evo} and introduce a scalar field $\tau$ to capture the local temperature, see Eq.~(5.9)\ecite{Crossley:2015evo}. This is an over-complete set, which they then constrain. 

Let us first see that the two sets of variables can be mapped onto each other; the choice made in \cite{Crossley:2015evo} can be obtained from \cite{Haehl:2015pja} by picking a suitable gauge for the reference fields. The argument is a mild generalization of the one given in Section 7.5 of \cite{Haehl:2015pja}. In particular, Eq.~(7.23)\ecite{Haehl:2015pja} picks a gauge where the reference thermal twist is made to vanish and the reference thermal vector is fixed  to a constant vector along the time direction. With this choice, the residual symmetries are as in Eq.~(7.24)\ecite{Haehl:2015pja}, viz., transverse spatial diffeomorphisms, chemical shift, and a general spacetime dependent time reparameterization (thermal shift).  To connect with \cite{Crossley:2015evo} we  want to reduce the time reparameterization part to depend only on the spatial coordinates. This is achieved by demanding that we fix ${\mathbb \bbbeta}^{a=I} = 0$, but pick ${\mathbb \bbbeta}^{a=0}$ to be an arbitrary function.\footnote{ We note  that the more recent paper \cite{Glorioso:2016gsa} GL switch to the thermal vector and twist variables.}

\subsection{Hydrodynamics as a theory of non-linear response}

 Both groups realize that usual hydrodynamics does not retain information about all the correlation functions of the system, but only cares about response correlators. The most general response functions are best described by the Schwinger-Keldysh formalism, where one is able to compute correlation functions with a specified amount of time-ordering.\footnote{ It is perhaps better to phrase the Schwinger-Keldysh formalism as being the right framework to compute singly out-of-time order correlation functions, amongst which appear the response functions of interest. Fluid dynamics correlators which are more out-of-time-order than Schwinger-Keldysh correlators are largely unexplored; though see the recent works of \cite{Aleiner:2016eni,Haehl:2016pec,Haehl:2017qfl}. } 

The microscopic computation of correlation functions in the Schwinger-Keldysh formalism entails a doubling of degrees of freedom as is well known (see \cite{Chou:1984es} for a review). This perspective was first applied to hydrodynamics in the context of anomalies in \cite{Haehl:2013hoa}. An inspirational paper that was helpful in further development was \cite{Kovtun:2014hpa} which in turn took inspiration from the Martin-Sigga-Rose construction \cite{Martin:1973zz}. The naive issues with directly upgrading it to the hydrodynamic effective field theory were explained in \cite{Haehl:2015pja}, which then prompted a more comprehensive analysis of the symmetry structure of the Schwinger-Keldysh formalism. We will describe the latter below, noting for now that the hydrodynamic description should involve both the classical macroscopic variables $X^\mu$ introduced above, as well as the fluctuation fields $\tilde{X}^\mu$, a feature emphasized in \cite{Haehl:2015foa,Crossley:2015evo,Haehl:2015uoc}.

The hydrodynamic currents are given in the Schwinger-Keldysh formalism as the variation of the effective action with respect to the advanced/difference fields (called `Adv' in \cite{Haehl:2015foa,Haehl:2015uoc,Haehl:2016pec} in thermal states, while \cite{Crossley:2015evo} refer to them as `a' fields.) The expression for the currents appears in Eq.~(6.6)\ecite{Haehl:2013hoa}, Eq.~(13.12)\ecite{Haehl:2015pja} and Eq.~(3.12)\ecite{Crossley:2015evo}, which can all be seen to be in agreement.

\subsection{BRST symmetries of the Schwinger-Keldysh formalism}

 The BRST symmetries are the first point of divergence between the  formalism of HLR and that of CGL.  Both groups use as their starting point the fact that correlation functions of difference operators vanish in the Schwinger-Keldysh construction. This is explained in Section 2.2 of \cite{Haehl:2015foa}, Eq.~(2.15)\ecite{Crossley:2015evo} and Eq.~(2.2)\ecite{Haehl:2015uoc}, and in Section I.D of \cite{Crooks:1998uq}, Eq.~(1.41)\ecite{Crossley:2015evo}.  Note that this does not capture all the structure that is inherent in Schwinger-Keldysh construction. For one, there is the largest time equation for the difference operators
which is not captured by demanding that the correlation functions of difference operators
vanish. These statements have been reviewed in greater detail more recently in \cite{Haehl:2016pec}.

HLR make the argument that the field redefinition redundancies owing to the Schwinger-Keldysh doubling can be understood as a pair of BRST charges  $\{\QSK, \QSKb\} $ with the difference/advanced operator being BRST exact with respect to both. This can be viewed in terms of two directions of descent from the average operator as first depicted in Eq.~(2.8)\ecite{Haehl:2015foa}:
\begin{equation}
\begin{tikzcd}
&\SKAv{O} \arrow{ld}{\QSK} \arrow{rd}[below]{\QSKb}    &   \\
\SKG{O}\arrow{rd}{\QSKb} & & \SKGb{O} \arrow{ld}[above]{\!\!\!\!\!\!\!\!\!\!\!\!\!\!-\QSK}\\
&   \SKDif{O} &
\end{tikzcd}
\label{eq:qskactionAvDif}
\end{equation}
As explained in Section 7.5 of \cite{Haehl:2016pec}, if we view $\QSK$ as a BRST charge for field redefinitions, then $\QSKb$ is the corresponding anti-BRST charge. In other words, we choose to work in the framework where both the symmetries are manifest (which is true at the level of the effective action). HLR also advocate a superspace built on two Grassmann odd variables, demanding a quadrupling of the operator algebra, as an efficient way to encode the BRST action (the $\QSK$ and $\QSKb$ act as superderivations). As Vafa-Witten \cite{Vafa:1994tf} explain in their Section 2.3,  the correct way to count the solutions without signs involves  working with \emph{two} supercharges, in a formalism which later came to be known as $\mathcal{N}_\smallT =2$ topological supersymmetry \cite{Dijkgraaf:1996tz}. As explained in \cite{Vafa:1994tf}, the formalism only one topological supersymmetry counts solutions with signs and  hence computes something more akin to an index.

On the other hand CGL decide to impose a single BRST charge $\delta$ to enforce the vanishing of Dif/Adv (`a' in their notation) operator correlators. This indeed suffices to ensure the topological limit.

However, there is more to be said. At a practical level, the most efficient way to implement the Schwinger-Keldysh BRST symmetry is by working in superspace which makes it manifest, as was first appreciated in \cite{Haehl:2015foa,Haehl:2015uoc}. Let us now describe this superspace more explicitly. Even the presence of a single BRST charge still demands that we quadruple the operator content. While the Av/Ret and correspondingly Dif/Adv fields are Grassmann even, the BRST action demands that they have corresponding Grassmann odd partners (the ghost fields). HLR encode this in a single superfield, cf., Eq.~(2.3)\ecite{Haehl:2015uoc} or Eq.~(6.19)\ecite{Haehl:2016pec}:
\begin{equation}
\SF{\Op{O}} = \SKRet{O} + \thb \, \SKG{O} + \theta\, \SKGb{O} + \thb\,\theta\, \SKAdv{O}\,,
\label{eq:hlrSF}
\end{equation}	
such that $\QSK$, $\QSKb$ act as $\partial_\thb$, $\partial_\theta$, respectively. 
Extrapolating the analysis of CGL to superspace, it will involve a single super-direction (parameterized say by $\thb$) and one would have two superfields with opposite Grassmann statistics. More specifically, we can reinterpret CGL's Eq.~(1.45)\ecite{Crossley:2015evo} in superspace as 
\begin{equation}
\SF{\Op{O}}_a = \SKRet{O} + \thb \, \SKG{O} \,, \qquad \SF{\Op{O}}_d = \SKGb{O} + \thb \, \SKAdv{O} \,.
\label{eq:mitSF}
\end{equation}	

We now wish to give a few arguments in favour of having two supercharges instead of one, and correspondingly a superspace representation as in \eqref{eq:hlrSF} instead of \eqref{eq:mitSF}. 

There is a natural pairing in the Schwinger-Keldysh formalism between the forward (R) and backward contours (L). We find it reasonable to encode both into a single superfield. 
Moreover,  the effective action  of CGL, Eq. (1.53)\ecite{Crossley:2015evo} has not only the BRST symmetry displayed in Eq.(1.54)\ecite{Crossley:2015evo}, but also an anti-BRST symmetry, obtained by exchanging their ghosts and anti-ghosts, and reversing the sign of their Adv (`a') field. 


The  Vafa-Witten argument for counting without signs via $\mathcal{N}_\smallT =2$ topological supersymmetry also demands two supercharges \cite{Vafa:1994tf}. Apart from this, perhaps the strongest argument in favour of two Schwinger-Keldysh supercharges is the following: effective actions for dissipative hydrodynamics have been constructed in \cite{Crossley:2015evo} and \cite{Haehl:2015uoc}. As we observed in \cite{Haehl:2015uoc}, the action does posses both symmetries without violating any phenomenological principles that we are aware of. The logic of standard effective field theory suggests that if the desired effective action has a symmetry, then there is no question about its existence, but merely one about its motivation from microscopics. The latter task was our undertaking in \cite{Haehl:2015foa,Haehl:2016pec,Haehl:2016uah}.\footnote{ The only caveat here is that the complete action for eightfold hydrodynamic transport has not been written down yet. However, given our knowledge about adiabatic Lagrangians from \cite{Haehl:2014zda}, and since the dissipative class is the one that is most sensitive to symmetry arguments, we would be rather surprised if other classes will eventually be inconsistent with $\QSKb$ symmetry.}

Finally, we note some of the advantages of working with a larger symmetry which unsurprisingly constrains more correlators and reduces many of the ambiguities in the ghost dynamics. HLR have shown that correlation functions involving ghosts can be fixed (up to certain ambiguities) in terms of the Av/Dif or Ret/Adv correlators. The ambiguities can be systematically classified as we explain in Section 9 of \cite{Haehl:2016pec}. This is not without its subtleties (suitable background ghost insertions are needed) and two supercharges were crucial to control the ambiguities. It seems to us that an analogous analysis with only one supercharge would greatly multiply ambiguities
and  would be practically intractable. We see no clear countervailing advantage in ignoring the symmetry and adopting a less symmetric formalism.

\subsection{KMS symmetries}

The second major difference in the two formalisms relies on how the KMS condition which arises in thermal (Gibbsian) states is implemented. We will later see that when the dust settles the two approaches will be quite similar.

HLR note that the KMS condition implies a set of thermal sum rules, which are over and above the Schwinger-Keldysh sum rules. These were clearly derived for instance in \cite{Weldon:2005nr}, and this argument was reviewed in Section (2.3) of \cite{Haehl:2015foa} (see the discussion following Eq.~(2.13)\ecite{Haehl:2015foa}), and  in Eq.~(4.20)\ecite{Haehl:2016pec}. 

In order to enable the reader to draw a quick conclusion it was pointed out in \cite{Haehl:2015foa,Haehl:2016pec,Haehl:2016uah} that one can ensure this thermal Ward identity by positing a new set of KMS symmetry generators $\{\QKMS, \QKMSb\}$. We realize that this leads to an interpretation where we have four BRST charges when the initial state is the thermal density matrix. However, as already noted in Section 2.4 of \cite{Haehl:2015foa} it is erroneous to view all four of these Grassmann odd generators as BRST differentials that pick out a cohomology. The SK-charges are true BRST differentials while the KMS-charges are interior contractions. Loosely speaking, the SK-charges are Weil differentials in the language of equivariant cohomology, while the KMS-charges are two particular contractions. The precise interpretation is explained in detail in Section 6 of \cite{Haehl:2016uah}.

The implementation of KMS symmetry by the CGL is different. As is common practice they combine the action of KMS and {\sf CPT}, to obtain a statement in terms of the generating functional, cf.,  Eq.~(1.69)\ecite{Crossley:2015evo}.\footnote{ A discussion disentangling various issues can be found in \cite{Sieberer:2015hba,Sieberer:2015aa}.} Since the action involves conjugation with {\sf CPT}, the resulting symmetry is viewed as a ${\mathbb Z}_2$ discrete symmetry. In Section 
I.G they posit that a useful way to encode the KMS conditions at low energies is in terms of an emergent fermionic symmetry Eq.~(1.75)\ecite{Crossley:2015evo}. They are clear that the symmetry operates only on the low energy action and make no statements at the microscopic level.

HLR have a different take on the low energy symmetries: they conjecture that the macroscopic manifestation of KMS is $\UT$ discovered in \cite{Haehl:2014zda,Haehl:2015pja}.  As indicated earlier, {\sf CPT} is implemented as a  ${\mathbb Z}_2$ involution (an R-parity acting anti-unitarily on the superspace), which  furthermore is spontaneously broken in the fluid phase. This is explained in 
Eq.~(2.5)\ecite{Haehl:2015uoc} and elaborated upon in  Eqs.~(8.6)\ecite{Haehl:2016pec} and (8.7)\ecite{Haehl:2016pec}. We will be slightly more detailed in Section \ref{sec:KMS} below. 

\subsection{The superalgebra underlying hydrodynamics}

 While the underlying intuition for dealing with thermal states in the two approaches in different, there is a surprising amount of concordance in the final superalgebra. CGL state that the superalgebra which constrains the low energy theory is given by Eq.~(1.77)\ecite{Crossley:2015evo}, reproduced here for convenience:
\begin{equation}
\delta^2 = 0 \,, \qquad {\bar \delta}^2 =0\,, \qquad [\delta,{\bar \delta}] =2\, {\bar \epsilon}\, \epsilon
\, \tanh\left(\frac{i}{2}\, \beta \partial_t\right) \approx    {\bar \epsilon} \, \epsilon \, i \beta \partial_t \,.
\label{eq:cglsusy}
\end{equation}	
It is argued that ${\bar \delta}$ is emergent at low energies and arises from the KMS condition. We will now demonstrate that at high temperatures this algebra can be precisely identified with a gauge fixed version of the superalgebra of HLR, first written down in \cite{Haehl:2015foa}. The high temperature limit of this algebra has also been noted earlier in the context of Langevin dynamics; cf.,  \cite{Mallick:2010su}.

The story in HLR is a-priori more detailed as necessitated  by a full implementation of the $\mathcal{N}_\smallT=2$ balanced equivariant formalism of Vafa-Witten \cite{Vafa:1994tf} and Digkgraaf-Moore \cite{Dijkgraaf:1996tz}. Their SK-KMS  superalgebra is characterized by a total of six generators denoted as $\{\QSK, \QSKb, \QKMS, \QKMSb, \Qzero, \Qbeta\}$. It is first argued that the low energy (equivalently $\beta \to0$ limit) is characterized by an emergent thermal translational symmetry $\UT$, as phenomenologically motivated by \cite{Haehl:2014zda,Haehl:2015pja}. In this limit, the six generators are argued to generate an $\mathcal{N}_\smallT =2$ equivariant cohomology algebra with the gauge symmetry being thermal translations. There are several ways to simplify the discussion, which are explained in some detail in Appendix A of \cite{Haehl:2015foa} and this formalism is heavily employed in its superspace version in \cite{Haehl:2015uoc}. It is useful to directly invoke the Cartan model for equivariant cohomology, undertake a partial gauge-fixing to the Wess-Zumino gauge, and show that the theory can be described by working with two gauge covariant Cartan charges $\Q$ and $\Qb$. Being gauge covariant, they are not nilpotent, but rather square to a gauge transformation, which in the present case is captured by $\Qbeta$. One finds by manipulating Eq.~(A.14)\ecite{Haehl:2015foa}:\footnote{ The relevant arguments are explained in some detail in Sections 5 and 6 of \cite{Haehl:2016uah}.}
\begin{equation}
\Q^2 = (\Fs_{\thb \thb} |_{_{\thb=\theta=0}} )\, \Qbeta \,, \qquad \Qb^2 = (\Fs_{\theta \theta} |_{_{\thb=\theta=0}}) \, \Qbeta \,, \qquad 
\gradcomm{\Q}{\Qb} = (\Fs_{\theta\thb}|_{_{\thb=\theta=0}}) \, \Qbeta
\label{eq:hlrQ1}
\end{equation}	
As a differential operator $\Qbeta$ is realized as the operation $\delKMS= -i(1-e^{-i\, \deltaB}) \approx \deltaB$, which implements a translation in the thermal direction, introduced in \cite{Haehl:2015foa}. The super-field strength $\Fs_{IJ}$ is the one associated with the gauge field of $\UT$. 

We can now compare the two algebras: suppose we identify  $\Q$ and $\Qb$ of HLR in \cite{Haehl:2015foa,Haehl:2016uah} with ${\bar \delta}$ and $\delta$, respectively of 
\cite{Crossley:2015evo}. The disadvantage of the equivariant cohomology presentation is that there are a whole slew of  extra ghost fields (the Vafa-Witten quintet of \cite{Haehl:2016uah}). It has been argued in \cite{Haehl:2015foa,Haehl:2016uah} that most of the covariant ghost of ghost fields which include $\Fs_{\thb \thb} | $ and $\Fs_{\theta \theta} | $ can be gauge fixed to zero, insofar as the macroscopic dynamics of the hydrodynamic fields is concerned. The only field that plays a non-trivial role is the ghost number zero field  $\Fs_{\theta\thb}| $, which picks up a non-trivial vacuum expectation value in the thermal state, owing to spontaneous {\sf CPT} symmetry breaking in a dissipative system. Note that it is an important ingredient of our approach to recognize the emergent second law and arrow of time in the fluid system as a consequence of the spontaneous breaking of the microscopic {\sf CPT} symmetry (see \cite{Gaspard:2012la,Gaspard:2013vl} for a detailed discussion). The $\UT$ symmetry provides a dynamical mechanism for this: we choose the {\sf CPT} breaking value $\langle\Fs_{\theta\thb}|\rangle = -i$ for the order parameter of dissipation.\footnote{ See, e.g., \cite{Haehl:2015uoc} for a motivation of this choice.}  With this understanding, we can simplify \eqref{eq:hlrQ1} to\footnote{ There are algebraic subtleties with the interpretation of signs, which are explained in \cite{Haehl:2016uah}.}
\begin{equation}
\Q^2 =0\,, \qquad \Qb^2 = 0\,, \qquad 
\gradcomm{\Q}{\Qb} = -i \,\Qbeta  \mapsto  i\lieD_\Kref
\label{eq:hlrQ2}
\end{equation}	
In general $\lieD_\Kref$  Lie drags operators along the thermal circle\footnote{ The operator $\lieD_\Kref$ refers to the Lie drag operation on worldvolume along the background vector $\Kref^a$.}, but on scalars it acts as $\Kref^a \partial_a$. In the static gauge, where $\Kref^{a=0} = \beta, \ \Kref^{a=I}=0$ this is indeed $\beta\, \partial_t$, which then gives a precise correspondence between the two algebras \eqref{eq:cglsusy} and \eqref{eq:hlrQ2}.  

To put it in a nutshell, despite the differing motivations, the superalgebras used by the two groups to constrain hydrodynamic effective actions is the same in the high temperature limit. The main distinction is that \eqref{eq:cglsusy} extends to beyond the high temperature limit and thus is aware of the detailed quantum statistics.\footnote{ The high temperature limit, as the reader can immediately appreciate is equivalent to the classical limit, since the quantum statistical distributions, Bose-Einstein or Fermi-Dirac, degenerate to the classical Maxwell-Boltzmann distribution.} HLR have not made a conjecture about the quantum  algebra, though it is easy to speculate that the structure is closely connected to that obtained by exponentiating the $\UT$ algebra introduced in \cite{Haehl:2015pja} to figure out the finite action of  $\UT$ as a group. While we have not been as yet able to prove it, we would be willing to speculate that the requisite group is the Virasoro-Bott group obtained by exponentiating the central extension of $\text{Vect}(\bf S^1)$ (the algebra associated with the group of diffeomorphisms on the circle $\text{Diff}({\bf S}^1)$). 

The basic common lesson learnt from the two approaches for constructing effective field theories in local thermal equilibrium then crystallizes as follows: the dynamics should be constrained by a superalgebra of the forms given above. We believe that the most efficient way to do this is by working with a superfield representation of the quadrupled space of low energy degrees of freedom, similar to the microscopic version given in \eqref{eq:hlrSF}. The difference to the latter is that KMS condition now needs to be implemented on top of generic SK unitarity. This translates to a covariantization of the superfields and the super-derivatives with respect to thermal translations. The mathematical framework for this was described in detail in \cite{Haehl:2016uah}. We note that in the high temperature regime the representation on fields of the derivations $\{\delta,\bar{\delta}\}$ of CGL is precisely the same as the covariant super-derivatives $\{\Dsf_\thb,\Dsf_\theta\}$ implementing $\{\Q,\Qb\}$ in the HLR formalism, for example (2.12)\ecite{Haehl:2015uoc}. Again, this statement holds after gauge fixing to zero all $\UT$ gauge field components apart from the flux condensate $\langle\Fs_{\theta\thb}|\rangle$.

\section{Is there a KMS $\UT$ symmetry?}
\label{sec:KMS}

Having argued that the algebras are for all intents and purposes the same, let us ask the only remaining question. Is there really an equivariant gauge symmetry, enshrined in the KMS $\UT$ symmetry? The arguments of HLR for this range from their initial discussion in  \cite{Haehl:2014zda,Haehl:2015pja} (see Section 15 of the latter, especially Eqs.~(15.2)\ecite{Haehl:2015foa}
and (15.3)\ecite{Haehl:2015foa}) and it is crucially embodied in their basic philosophy espoused in  \cite{Haehl:2015foa}. 
The formal arguments in favour of this are explained in some detail in \cite{Haehl:2016uah}, but the real  proof of its viability is in its constraining of the hydrodynamic effective action \cite{Haehl:2015uoc}. 
As explained after \eqref{eq:LTact}, in the adiabatic sector of hydrodynamics the emergence of a $\UT$ symmetry is a phenomenological fact. The question we want to address here is whether or not it should be gauged, and what is its fate in the dissipative sector of hydrodynamics. 

At first sight, the action for dissipation presented in \cite{Crossley:2015evo} appears to be rather different from that in \cite{Haehl:2015uoc}. Of course, much of this is cosmetic owing to a different choice of variables etc. Fortunately, rather than work out the precise map, we can take the recent attempt to understand the second law and entropy current by GL \cite{Glorioso:2016gsa} as our starting point. 

\subsection{Of accidental symmetries}
\label{sec:symms}

Let us start with an examination of the hydrodynamic effective action presented in Appendix D.2 of \cite{Glorioso:2016gsa}. They examine the transformation of the `a' type field (what we call $\tilde{X}^\mu$ here) under the KMS symmetry; in particular, the symmetry they impose, Eq.~(D23)\ecite{Glorioso:2016gsa}, reads in our language:
\begin{equation}
\partial_a \tilde{X}_\mu (-\sigma) \rightarrow  \partial_a \tilde{X}_\mu(\sigma) + i\, \partial_a \Kref_\mu(\sigma) \,.
\label{eq:Xtgt}
\end{equation}	
Consider repackaging $\tilde{X}_\mu$ into an  auxiliary object 
$\tilde{g}_{\mu\nu}$ via $ \partial_\mu \tilde{X}_\nu + \partial_\nu \tilde{X}_\mu  = \tilde{g}_{\mu\nu} $, which is pretty much enforced by demanding the correct derivative counting and having a symmetric 2-tensor, and physically motivated by the considerations of Section 15 of \cite{Haehl:2015pja}. In fact, as explained there $\tilde{g}_{\mu\nu}$ is designed for precisely the purpose of capturing the hydrodynamical `a'-modes. With this identification of $\tilde{X}_\mu$, the KMS transformation \eqref{eq:Xtgt} translates into a statement for transformation of 
$\tilde{g}_{\mu\nu}$:
\begin{equation}
 \tilde{g}_{\mu\nu} (-x) \rightarrow  \tilde{g}_{\mu\nu}(x) + i \lieD_\Kref g_{\mu\nu}(x) 
 \label{eq:tgut}
\end{equation}	
where $g_{\mu\nu}$ is the background metric and $\lieD_\Kref$ denotes Lie derivative as above. Up to the sign flip of the coordinates in \eqref{eq:Xtgt} this is in fact precisely the $\UT$ part of Eq. (15.4)\ecite{Haehl:2015pja}  with the particular identification of $\UT$ gauge parameters 
$ \bar{\xi}^\mu=0 ,\  \bar{\Lambda}^{({\sf T})} = i $.  Thus, somewhat crucially \cite{Glorioso:2016gsa} employ a close relative of the $\UT$ transformation. They refer to it as a `dynamical KMS transformation' and as we will see momentarily it can be thought of as a discrete version of HLR's $\UT$ combined with standard {\sf CPT}. 

Let us see this a different way: consider Eqs.~(5.1)\ecite{Haehl:2015uoc}  which display the $\UT$ transformation on the worldvolume superfields. Working out the transformation of the $\thb\theta$-component, we find Eq.~(5.2)\ecite{Haehl:2015uoc}, where the pullback metric ${\sf g}_{ab}$'s Adv/Dif-partner is $\tilde{{\sf g}}_{ab}$, and its transformation is exactly as indicated in \eqref{eq:tgut}.
 As explained in some detail in \cite{Haehl:2015uoc} the $\thb\theta$-component of a given superfield transforms through a shift involving the bottom component (which is the Ret/Av or `r' component). 
 
We can summarize the philosophies regarding the implementation of KMS condition as follows: 
\begin{itemize} 
\item CGL impose a `dynamical KMS transformation' as in \eqref{eq:Xtgt}, which is a suitable discrete combination of a KMS time translation with {\it PT} reversal. 
\item HLR impose standard {\sf CPT} and a continuous $\UT$ symmetry. There is however an interplay between these two symmetries: {\sf CPT} is spontaneously broken in the dissipative fluid phase and the order parameter for this symmetry breaking is argued to be a particular flux condensate of the $\UT$ field strength component, viz., $\langle\Fs_{\theta \thb} |\rangle$.\footnote{ The phenomenology of such a symmetry breaking is briefly discussed in Section 5 of \cite{Haehl:2015uoc}.} 
\end{itemize}

We note that the second approach also entails a discrete $\mathbb{Z}_2$ symmetry, which is a combination of $\UT$ and {\sf CPT}:\footnote{ We thank Hong Liu for a discussion on this point, which helped us to see this connection more clearly.}
in order to derive the second law, HLR  employ a  $\mathbb{Z}_2$ transformation, which can be described as a particular $\UT$ transformation combined with {\sf CPT}, see Section 7.5\ecite{Haehl:2016uah} for a detailed discussion that also applies to hydrodynamics \cite{Haehl:2015uoc}. For instance, one of the $\thb\theta$-components of the superfield transformations (7.54)\ecite{Haehl:2016uah} with (7.53)\ecite{Haehl:2016uah} made explicit, reads as
\begin{equation}\label{eq:KMStrf}
  \tilde{X}(t) \;\mapsto\; \tilde{X}(-t) - (\Fs_{\theta \thb} |)\, \delKMS X(-t) \,.
\end{equation}
To account for dissipation in the effective theory we choose a {\sf CPT} breaking expectation value $\langle \Fs_{\theta \thb} |\rangle = -i$ and find precise agreement with GL's dynamical KMS symmetry \eqref{eq:Xtgt} or (2.12)\ecite{Glorioso:2016gsa}.
It is gratifying to see that GL's guiding principle for constructing an entropy current consistent with the second law (i.e., the dynamical KMS $\mathbb{Z}_2$ symmetry) can be put in precise correspondence with the transformation that can be used within the HLR formalism to prove the second law using the mechanism of spontaneous {\sf CPT} breaking in the language of $\UT$ emergent gauge theory.\footnote{ Note that in the HLR construction, while \eqref{eq:KMStrf} can be used to derive the second law, it is perhaps more natural to consider their implementation of {\sf CPT} as the discrete symmetry whose involution with $\UT$ gives the desired (equivalent) results. This has been done in version 2 of \cite{Haehl:2016uah}, see eq.\ (7.46)\ecite{Haehl:2016uah}.}


There is a final similarity of the formalisms that is worth indicating. It is appreciated by CGL in \cite{Crossley:2015evo,Glorioso:2016gsa} that there is in fact an additional continuous symmetry (at least at linear order in `a' type fields), cf., Eq.~(3.15)\ecite{Glorioso:2016gsa}. While its origin lies in the discrete transformation \eqref{eq:Xtgt}, it is clear that the transformation  Eq.~(3.15)\ecite{Glorioso:2016gsa}  corresponds accurately with the  $\UT$ symmetry of HLR, c.f., Eq.~(5.2)\ecite{Haehl:2015uoc}. Note that the hydrodynamic effective action with background $\UT$ flux  does not manifest the full symmetry. We believe this to be responsible for CGL and GL noting that there is a continuous accidental symmetry at ideal fluid level, but not at higher orders. The analysis of HLR goes beyond the appreciation of this symmetry in three respects: 
\begin{enumerate}
\item[$(i)$] It shows that the symmetry extends beyond ideal fluids to 7 out of 8 classes of hydrodynamic transport, and gives a dynamical symmetry breaking mechanism to explain dissipation. 
\item[$(ii)$] HLR introduce a gauge field associated with this symmetry. In our eyes this has many useful consequences, the most prominent one being that the characterization of entropy as a Noether current becomes both manifest and easy to implement. 
\item[$(iii)$] We argue that the $\UT$ symmetry is in fact gauged. Arguments in favour of gauging and the issues associated with it were already noted in Appendix B of \cite{Haehl:2015foa}. Perhaps the simplest argument for gauging is the use of the machinery of extended equivariant cohomology. To date, we are not aware of any issue with the symmetry being gauged, though not all details have as yet been worked out. As indicated in the aforementioned references, there is likely to be a non-trivial BF type topological theory for the $\UT$ gauge field. 
\end{enumerate}

\subsection{Of entropy current and KMS shifts}

The cleanest identification of the accidental symmetry postulated in \cite{Glorioso:2016gsa} with HLR's $\UT$ follows from GL's identification of the  entropy current as the Noether current associated with a continuous symmetry that shifts `a'-fields by `r'-fields. This is the primary reason behind the introduction of  $\UT$ in \cite{Haehl:2015pja}. A crucial insight of that work was indeed that that entropy current is the Noether current for $\UT$. The analysis there was inspired by adiabaticity equation mentioned earlier, Eq.~(2.12)\ecite{Haehl:2015pja}, and how the free energy current in the Landau-Ginzburg sector of hydrodynamics is a Noether current associated with thermal translations, see (6.18)\ecite{Haehl:2015pja}.\footnote{ Historically,  S. Bhattacharyya's papers  \cite{Bhattacharyya:2013lha,Bhattacharyya:2014bha} on hydrostatic partition functions were crucial for the development of these ideas by HLR.}

In addition to the action on fields, one would hope that the KMS symmetry action of GL and  $\UT$ transformation of HLR would produce the same effect on the effective action. Indeed, this is true, as can be seen by comparing Eq.~(3.3)\ecite{Glorioso:2016gsa} with the $\UT$ transformation probing spontaneous {\sf CPT} breaking in a dissipative fluid, Eqs. (5.4)\ecite{Haehl:2015uoc} and (5.5)\ecite{Haehl:2015uoc}. To be explicit, the operator $W^\mu$\ecite{Glorioso:2016gsa} is what we call the (negative of) the free energy current $N^a$\ecite{Haehl:2015uoc}, defined in \eqref{eq:Ncurrent}. The agreement appears to be exact in both the total derivative terms, and in the terms obtained by expanding the effective action to quadratic order in the Adv or `a' type fields (modulo the fact that \cite{Glorioso:2016gsa} do not have the $\UT$ gauge field).

One can take the similarities between the analyses further. Eq.~(3.12)\ecite{Glorioso:2016gsa} is a slight generalization of HLR's adiabaticity equation,
 as can be seen by comparing it to the free energy version given in Eq.~(2.21)\ecite{Haehl:2015pja} (we recall that this has been the guiding principle behind the analysis):\footnote{ We drop here contributions from anomalies and only account for Abelian flavour charges for simplicity.}
\begin{equation}\label{eq:AdiabaticityG}
\begin{split}
\nabla_\sigma N^\sigma &=
\half  T^{\mu\nu} \lieD_\Kref  g_{\mu\nu} + J^\mu \cdot \left( \lieD_\Kref  A_\mu + \partial_\mu \LambdaB  \right)  + \Delta \,,
\end{split}
\end{equation}
where $\Delta$ denotes the total (off-shell) entropy production. 
The structural form of the dissipative class is governed by a four tensor, $T^{ab}_{diss} = \frac{1}{2}{\bm \eta}^{(ab)(cd)} \lieD_\Kbeta g_{cd}$, which defines a positive definite inner product on maps from the space of symmetric two-tensors to symmetric two-tensors.\footnote{ We are not writing the most general form of $\Delta$ here. See \cite{Haehl:2015pja} for details.} The dissipation then takes the form
\begin{equation}
\Delta = \frac{1}{4} {\bm \eta}^{(ab)(cd)} \lieD_\Kref  g_{ab} \, \lieD_\Kref g_{cd} \,.
\end{equation}
This has been uncovered in Eq.~(5.7)\ecite{Haehl:2015pja}, first derived in this form from an effective action in \cite{Haehl:2015uoc}, and now been confirmed by GL in Eq.~(D26)\ecite{Glorioso:2016gsa} -- all in mutual agreement.

\subsection{Of dissipative hydrodynamic action}
\label{sec:}

Thus far we have seen that the underlying symmetries can be shown to match and the basic results of \cite{Glorioso:2016gsa} relating to the entropy current can be matched to the analysis of \cite{Haehl:2015pja}. Let us then finally check that the effective actions agree, as it is clear they must at this point.

The hydrodynamic effective actions which we compare are the ones given by GL in Eq.~(D24)\ecite{Glorioso:2016gsa} and by us in Eq.~(4.4)\ecite{Haehl:2015uoc}. To guide the eye, let us reproduce the dissipative part of the latter here: 
\begin{align}
S_\text{wv} &= \frac{1}{4} \int  \frac{d^d\sigma \, \sqrt{-\gref}}{1+\Kref^e \Ascr_e} \;\bigg\{ \!
	 - \etaref^{(ab)(cd)}\, \left( i {\mathscr F}_{\theta\bar{\theta}}|, 
	\gref_{cd}\right)_{\Kref}\tilde{\gref}_{ab} 
	 +  i\,\etaref^{((ab)|(cd))} \, \tilde{\gref}_{ab} \, \tilde{\gref}_{cd}
	+ \ldots \bigg\}\,,
\label{eq:classLTa}
\end{align}
where Grassmann-odd directions of the worldvolume thermal superspace have already been integrated out, and $\Ascr_e$ denotes the $\UT$ gauge field in this context. A similar form for the action has already been advocated in \cite{Kovtun:2014hpa}.
We can almost immediately see that it is of the same form as Eq.~(D24)\ecite{Glorioso:2016gsa}:
\begin{itemize}
\item  By suitably symmetrizing their first term as described below \eqref {eq:Xtgt}, we can convert $\tilde{X}_\mu$ into $\tilde{{\sf g}}_{\mu\nu}$. This term then is indeed the one expect from the analysis of the eightfold Lagrangian \eqref{eq:LTact} (second term there).
We can be  more explicit: the first term of Eq.~(D24)\ecite{Glorioso:2016gsa}, or the second term in Eq.~(2.9)\ecite{Glorioso:2016gsa} are  the same as the second term of Eq.~(15.25)\ecite{Haehl:2015pja}, which was reproduced from the thermal equivariant cohomology perspective in Eq.~(4.4)\ecite{Haehl:2015uoc} (first term there). 
The only real difference between the action of \cite{Glorioso:2016gsa} and the Class $\LT$ Lagrangian are the additional couplings to the $\UT$ gauge field (and the fact that the said gauge field is present in the first place).

\item   The second term of  Eq.~(D24)\ecite{Glorioso:2016gsa}  is  needed to capture dissipation and  corresponds to the second term in \eqref{eq:classLTa}. Hydrodynamic dissipation is governed by ${\bm \eta}^{(ab)(cd)}$
which couples to a bilinear of the Adv fields $\tilde{{\sf g}}_{ab}\,\tilde{{\sf g}}_{cd}$. We can identify the dissipative tensor ${\bm \eta}^{(ab)(cd)}$\ecite{Haehl:2015uoc} with  $W^{\mu\nu,MN}$\ecite{Glorioso:2016gsa}.\footnote{ The tensor $W^{\mu\nu,MN}$\ecite{Glorioso:2016gsa} has two type of indices because it at the same time captures dissipation through flavour charges. Analogous parametrizations can be found in \S5 of \cite{Haehl:2015pja}.} The statements about entropy production Eq.~(D31)\ecite{Glorioso:2016gsa} are then identified with the previous expressions appearing in Eq.~(5.15)\ecite{Haehl:2015pja}, or Eq.~(5.8)\ecite{Haehl:2015uoc}. 
\item Concerning symmetries, in the dissipative phase (characterized by $\langle {\mathscr F}_{\theta\bar{\theta}}| \rangle \neq 0$) the continuous $\UT$ symmetry of the action \eqref{eq:classLTa} is not manifest anymore, as advertised. As reviewed in Section \ref{sec:symms}, there is a particular $\mathbb{Z}_2$ combination of $\UT$ transformation and {\sf CPT} which probes the breaking of time reversal invariance and can be identified with CGL's dynamical KMS symmetry. This is the symmetry which CGL use to derive \eqref{eq:classLTa}, whereas HLR derive the same action from a covariant superspace approach using $\UT$ equivariance. Both approaches invoke the same $\mathbb{Z}_2$ to derive the second law.
\end{itemize}

In summary, while the approaches are slightly different, the result (D24)\ecite{Glorioso:2016gsa} is exactly the same as the central result 
Eq.~(4.4)\ecite{Haehl:2015uoc} quoted above,  after setting $(i\Fs_{\theta\thb}|,{\sf g}_{ab})_\Kref =\lieD_\Kref {\sf g}_{ab}$ (as explained in \cite{Haehl:2015uoc}), and dropping the contribution from the $\UT$ gauge field $\Ascr_e$.

The physics lesson to be taken from both approaches which recover this result, is the structure of the couplings between Ret/Av- and Adv/Dif-fields in the SK doubled theory. Further, the factor $i$ in the second term in \eqref{eq:classLTa} links convergence of the path integral to positivity of the dissipative tensor (and hence of entropy production). Note that HLR's result (4.4)\ecite{Haehl:2015uoc} has more structure on top of the simple matching given above: the presence of the $\UT$ gauge field has useful consequences such as an immediate realization of dissipative entropy current as being obtainable from a variation with respect to $\Ascr_a$.

\section{Conclusion}

At a basic level it is gratifying to see the commonalities between the two distinct formalisms which have in the past few years been developed to tackle the problem of constructing hydrodynamic effective actions. There remains more to be done, but we hope that this short discussion serves to inform interested readers of the current status quo.
 
\acknowledgments
FH gratefully acknowledges support through a fellowship by the Simons Collaboration `It from Qubit'. RL gratefully acknowledges support from International Centre for Theoretical Sciences (ICTS), Tata institute of fundamental research, Bengaluru. RL would  also like to acknowledge his debt to the people of India for their steady and generous support to research in the basic sciences.


\providecommand{\href}[2]{#2}\begingroup\raggedright\endgroup

\end{document}